\documentclass[12pt]{article}
\usepackage[utf8]{inputenc}
\usepackage{graphicx}
\usepackage{amsmath}
\usepackage{amssymb}
\usepackage{amsfonts}
\usepackage{float}
\usepackage{listings}
\usepackage{academicons}
\DeclareUnicodeCharacter{0301}{\'}
\DeclareUnicodeCharacter{0308}{\"}

\usepackage{tikz,xcolor}

\definecolor{lime}{HTML}{A6CE39}
\DeclareRobustCommand{\orcidicon}{
	\begin{tikzpicture}
	\draw[lime, fill=lime] (0,0) 
	circle [radius=0.16] 
	node[white] {{\fontfamily{qag}\selectfont \tiny ID}};
	\draw[white, fill=white] (-0.0625,0.095) 
	circle [radius=0.007];
	\end{tikzpicture}
	\hspace{-2mm}
}
\foreach \x in {A, ..., Z}{\expandafter\xdef\csname orcid\x\endcsname{\noexpand\href{https://orcid.org/\csname orcidauthor\x\endcsname}
			{\noexpand\orcidicon}}
}
\usepackage[font=small,labelfont=bf]{caption}
\usepackage{color}
\usepackage{csquotes}
\usepackage[english]{babel}
\usepackage[affil-it]{authblk}
\usepackage[left=2.5cm,right=2.5cm, top=3cm, bottom=3cm]{geometry}
\usepackage{mathcomp}
\usepackage{breqn}
\usepackage{tabularx}
\usepackage{multirow}
\usepackage{makecell}
\usepackage{hyperref}
\usepackage{setspace}
\hypersetup{              
    pdftoolbar=true,        
    pdfmenubar=true,       
    pdffitwindow=false,     
    pdfstartview={FitH},
    pdfauthor={Parangam Goswami, Anshuman Baruah},     
    colorlinks=true,       
    linkcolor=mauve,         
    citecolor=dkgreen,       
}
\makeatletter
\newbox\abstract@box
\renewenvironment{abstract}
  {\global\setbox\abstract@box=\vbox\bgroup
     \hsize=\textwidth\linewidth=\textwidth
    \small
    \begin{center}%
    {\bfseries \abstractname\vspace{-.5em}\vspace{\z@}}%
    \end{center}%
    \quotation}
  {\endquotation\egroup}
\expandafter\def\expandafter\@maketitle\expandafter{\@maketitle
  \ifvoid\abstract@box\else\unvbox\abstract@box\if@twocolumn\vskip1.5em\fi\fi}
\makeatother
\providecommand{\keywords}[1]{\textbf{\textit{Keywords---}} #1}
\begin{document}
\definecolor{dkgreen}{rgb}{0,0.6,0}
\definecolor{gray}{rgb}{0.5,0.5,0.5}
\definecolor{mauve}{rgb}{0.58,0,0.82}

\lstset{frame=tb,
  	language=Matlab,
  	aboveskip=3mm,
  	belowskip=3mm,
  	showstringspaces=false,
  	columns=flexible,
  	basicstyle={\small\ttfamily},
  	numbers=none,
  	numberstyle=\tiny\color{gray},
	commentstyle=\color{dkgreen},
  	stringstyle=\color{mauve},
  	breaklines=true,
  	breakatwhitespace=true
  	tabsize=3
}
    \title{Traversable wormholes in $f(R)$ gravity sourced by a cloud of strings}
    \author[]{Parangam Goswami\thanks{E-mail: \textsf{parangam.goswami@aus.ac.in}}\orcidA{}}
    \author[]{Anshuman Baruah\thanks{E-mail: \textsf{anshuman.baruah@aus.ac.in}}\orcidB{}}
    \author[]{Atri Deshamukhya\thanks{Correspondence to: \textsf{atri.deshamukhya@gmail.com}}\orcidC{}}
    \affil{Department of Physics, Assam University, Cachar - 788011, Assam, India}
    \date{}
    
\begin{abstract}
Wormhole solutions in General Relativity (GR) require \textit{exotic} matter sources that violate the null energy condition (NEC), and it is well known that higher--order modifications of GR and some alternative matter sources can support wormholes. In this study, we explore the possibility of formulating traversable wormholes in $f(R)$ modified gravity, which is perhaps the most widely discussed modification of GR, with two approaches. First, to investigate the effects of geometrical constraints on the global characteristics, we gauge the $rr$--component of the metric tensor, and employ Pad\`{e} approximation to check whether a well--constrained \textit{shape function} can be formulated in this manner. We then derive the field equations with a background of string cloud, and numerically analyse the energy conditions, stability, and amount of exotic matter in this space--time. Next, as an alternative source in a simple $f(R)$ gravity model, we use the background cloud of strings to estimate the wormhole shape function, and analyse the relevant properties of the space--time. These results are then compared with those of wormholes threaded by normal matter in the simple $f(R)$ gravity model considered. The results demonstrate that wormholes with NEC violations are feasible; however, the wormhole space-times in the simple $f(R)$ gravity model are unstable. 
\\

\noindent \keywords{Wormhole, string cloud, modified gravity, energy conditions}
\end{abstract}

\maketitle

\tableofcontents

\section{Introduction}\label{sec1}
The governing equation of the theory of General Relativity (GR) is the Einstein field equation (EFE), and solutions to this set of coupled differential equations have been remarkably successful in accounting for phenomena ranging from black holes \cite{1916AbhKP1916..189S} to the evolution of the universe. Spherical symmetry is an important physically relevant constraint in solving the EFEs, and wormholes are an intriguing prospect among spherically symmetric solutions, explored since the early days of GR \cite{Flamm:1916,einstein1935particle}. Ellis \cite{ellis1973ether} and Bronnikov \cite{bronnikov1973scalar} independently reported the first traversable Lorentzian wormhole solution, and the geometric requirements were established in detail by Morris \& Thorne in 1988 \cite{Morris:1988cz}. Wormholes are exact solutions to the EFEs, and can be interpreted as bridges connecting two different asymptotically flat regions of space--time via a \emph{throat}. The topology of the wormhole interior is non-trivial, while the topology at the boundaries remains simple \cite{Visser:1995cc}. While theoretical as of now, wormholes are of crucial importance in fundamental physics, especially considering quantum entanglement \cite{epr} and quantum gravity. Moreover, it has recently been shown that wormholes may actually mimic astrophysical black holes in some observations \cite{nandi2017ring}. A fundamental problem in such space--times is that for the \emph{throat} to remain open for signal propagation, the behavior of the matter sources supporting such a space--time is \emph{exotic} in that the null energy condition (NEC) is violated \cite{Morris:1988cz}. Specifically, this implies that observers in an inertial frame measure negative energy densities at the throat, which is unphysical. Such behaviour can be avoided in modified gravity theories.

Despite its success, GR fails to explain cosmological phenomena such as late--time accelerated cosmic expansion \cite{perlmutter1999constraining,riess2001farthest} and the so called `inflationary epoch' \cite{linde1990particle}. Modified theories of gravity address these shortcomings \cite{starobinsky1980new,copeland2006dynamics,capozziello2011extended}, and the extra degrees of freedom in modified gravity theories also enables one to evade NEC violations in wormhole space--times. Moreover, wormhole solutions are an inherent feature of most modifications of GR. One of the simplest modifications to GR is $f(R)$ modified gravity, where the Ricci scalar $R$ in the Einstein--Hilbert action is replaced by some arbitrary function of it \cite{sotiriou2010f}. This simple modification to GR can lead to a host of models that can independently meet both cosmological and solar system tests \cite{guo2014solar}. It has been shown that the extra degrees of freedom in $f(R)$ gravity arising from higher order curvature terms may lead to scenarios where the matter content satisfies the NEC in wormhole space--times \cite{capozziello2015generalized, Baruah_2019}. Wormholes in the framework of $f(R)$ gravity have been studied extensively in different iterations of $f(R)$ gravity (for instance, in Refs. \cite{lobo2009wormhole, furey2004wormhole,godani2019non, godani2019traversable,azizi2013wormhole,mishra2020traversable,boehmer2012wormhole,mustafa2022traversable, baruah2022new, baruah2023non}).

Another inherent limitation of GR is that it provides only a classical description of gravity, and the theory is non--renormalisable at high energy (small length) scales. String theory \cite{mukhi2011string} is perhaps the strongest contender to a unified paradigm of gravity with the other fundamental forces, and it posits that the fundamental constituent of matter and energy are extended objects, instead of point--like ones. Precisely, the extended objects are considered as one--dimensional relativistic strings, and the interactions of strings on a classical level provide better models of several fundamental interactions \cite{kalb1974classical,letelier1977gauge,lund1976unified}. To this end, string clouds as a gravitational source have been studied extensively in literature. A general solution to the EFEs for a spherically symmetric cloud of strings was first reported in \cite{letelier1979clouds}, with an emphasis on the energy conditions. Properties of compact objects such as black holes in the background of a string cloud have been reported previously \cite{ghosh2014cloud,belhaj2022shadows,singh2020clouds}. Traversable wormholes in the background of a string cloud have been reported \cite{richarte2008traversable}, with an emphasis on the amount of exotic matter and stability of the wormhole configuration against radial perturbations. Recently, a detailed study of the properties of traversable wormholes surrounded by a string cloud in the framework of $f(R)$ gravity have been reported with analysis of the quasi normal modes (QNMs) of the wormhole solutions \cite{gogoi2023tideless}.

In this study, we investigate traversable wormholes in $f(R)$ gravity with two motivations: First, to investigate the effects of geometrical constraints, we gauge the $rr$--component of the metric tensor, and employ Pad\'{e} approximation to check whether a well--constrained \textit{shape function} can be formulated in this manner. We then derive the field equations with the background of a string cloud, and numerically analyse the energy conditions, stability, and amount of exotic matter in this space--time. Next, we use a background of string cloud to estimate the wormhole shape function in a simple $f(R)$ gravity model, with $f(R) = \alpha R^m - \beta R^n$ \cite{nojiri2008modified}, and analyse the relevant properties of the space--time as in the previous case. These results are then compared with those of wormholes threaded by normal matter in the same modified gravity model considered. 

The remainder of this manuscript is organised as follows. In Sec. \ref{sec2}, we discuss the traversable wormhole geometry. A novel shape function is proposed using Pad\'{e} approximation in Sec. \ref{subsec2_1}. In Sec. \ref{sec3}, we present the modified EFEs in the general framework of $f(R)$ gravity. In Sec. \ref{subsec3_1} we analyse the various energy conditions, the stability in terms of TOV equation, and the amount of exotic matter required to sustain traversible wormholes in the framework of $f(R)$ gravity with a background of string cloud. Next, in Sec. \ref{subsec3_2} we estimate the shape function in the considered form of $f(R) = \alpha R^m - \beta R^n$ gravity model \cite{nojiri2008modified} with a string cloud background and analyse the properties of the wormhole space-time. For a comparative analysis, in Sec. \ref{subsec3_3} we present wormhole solutions supported by ordinary matter for the $f(R)$ model \cite{nojiri2008modified} considered in Sec. \ref{subsec3_2} using our proposed shape function obtained by employing Pad\'{e} approximation in Sec. \ref{subsec2_1}.
Finally, we conclude the work with some remarks in Sec. \ref{sec4}. We adhere to use the natural system of units ($G=c=1$) throughout the work.

\section{Traversable wormholes}\label{sec2}
Morris \& Thorne \cite{Morris:1988cz} used the following metric ansatz to describe a static, spherically symmetric space-time
\begin{equation}
    \label{mtle}
     ds^2 = - e^{2 \Phi(r)} dt^2 + \frac{dr^2}{ 1 - \frac{b(r)}{r}} + r^2 {d{\theta}}^2 + r^2 sin^2 {\theta} d {\phi}^2
    \end{equation}
Eq. \eqref{mtle} is the line element of a traversable wormhole. The proper radial coordinate $l(r) = \int_{r_0}^r \frac{dr}{\sqrt{1-\frac{b}{r}}}$ should be well behaved throughout the space-time in order to avoid singularities. It imposes the constraint $\frac{b}{r} \leq 1$ at the throat. This space-time is a special case of the Ellis--Bronnikov \cite{ellis1973ether,bronnikov1973scalar} space-time described in terms of $l(r)$ as
\begin{equation}
    \label{Ellis_B_1}
     ds^2 = - dt^2 e^{2 \Phi (l)} + dl^2+ r^2(l) ({d{\theta}}^2 + r^2 sin^2 {\theta} d {\phi}^2)
    \end{equation}
where the throat is located at some minimum $r(l)$.\\
The metric function $\Phi(r)$ in Eq. \eqref{mtle} is known as the red-shift function, and the first co-efficient of the line element in Eq. \eqref{mtle} provides a measure of the gravitational red-shift. The topological configuration of the space-time is determined by the second coefficient of the line element in Eq. \eqref{mtle}, and the metric function $b(r)$ is known as the shape function. At some minimum value of the radial co-ordinate $r$, the throat of the wormhole is located at some arbitrary value $r_0$. A significant aspect of traversable wormholes is that the throat should not be surrounded by an event horizon. Horizons in spherically symmetric space-times are described by physically non-singular surfaces at $g_{00}=-e^{2\Phi} \rightarrow 0$, and this results in the constraint that throughout the space-time $\Phi(r)$ should be well defined. Moreover, the geometric constraints on the shape function $b(r)$ demanded by traversability are: (i) $b(r_o) = r_o$, (ii) $\frac{b(r) - b^{\prime} (r) r}{b^2} > 0$, (iii) $b^{\prime} (r_o) - 1 \leq 0$, (iv) $\frac{b(r)}{r} < 1, \forall r > r_o$, (v) $\frac{b(r)}{r} \rightarrow 0$ as $r \rightarrow \infty$, where prime denotes a derivative with respect to the radial co-ordinate $r$. The energy density $\rho$, radial pressure $p_{r}$, and transverse pressure $p_{t}$ of the matter sources are constrained by these conditions on the metric functions through the EFEs. Therefore, while constructing traversable wormhole configurations violations of the energy conditions appear owing to these constraints. Next, we propose a novel form the shape function and check the viability of the shape function by analysing the various constraints. In addition, in this work we consider tideless traversable wormhole solutions described by a constant red-shift function $\Phi'(r)=0$.

\subsection{A novel shape function}\label{subsec2_1}
We examine the following functional form as a probable shape function
\begin{align}
\label{nov_sf_ch_4}
b(r)=r_0 \left[\log \left(\frac{r}{r_0}\right) + \coth (r_0) \tanh (r) \right]^a
\end{align}
where, $r_0$ is the throat, and $a$ is a free parameter. The viability of the shape function $b(r)$ depends on several constraints as discussed in Sec \ref{sec2}. Analysing these constraints for the functional form of Eq. \eqref{nov_sf_ch_4},  insights can be obtained about the throat radius $r_0$, and the parameter $a$. However, it can be observed Eq. \eqref{nov_sf_ch_4}, does not satisfy the condition of $b(r_0) = r_0$, which is an important consequence of traversable wormholes, the other required conditions for the viability of the shape function, such as the asymptotic flatness, and the flaring out condition are also not satisfied. Thus, in order to have a plausible form of $b(r)$, the next step involves adopting the Pad\'{e} approximation for the functional form in Eq. \eqref{nov_sf_ch_4}, and to check if a viable form of $b(r)$ can be obtained that satisfies all the required constraints on a shape function. Use of rational expansions made by Pad\'{e} functions is common in existing literature \cite{capozziello2019extended,gruber2014cosmographic,zhou2016new,capozziello2021traversable}. The Pad\'{e} approximation is built on the Taylor series expansion. For a given function such as $f(z) = \sum_{i=1}^{\infty} c_{i} z^{i}$, expanding the series with the coefficients $c_{i}$, the $(n,m)$ Pad\'{e} approximant ratio is given as \cite{ pade1892representation},

\begin{align}
    P_{n,m} (z) = \frac{\sum_{k=0}^{n} a_{k} z^{k}}{1 + \sum_{\sigma=0}^{m} b_{\sigma} z^{\sigma}}
\end{align}

The most simple forms of Pad\'{e} approximant are of the orders $(1,0)$ \& $(0,1)$. Thus, considering the $(1,0)$ approximation and expanding Eq. \eqref{nov_sf_ch_4} about the throat radius $r_0$, the shape function is
\begin{align}
\label{main_sf_string}
b(r) = r_0 - a (r_0 - r) \left[1 + r_0~ \text{cosech}(r_0) \text{sech}(r_0) \right]
\end{align}
With the form as described in Eq. \eqref{main_sf_string}, the various conditions required for the viability are analysed and it becomes evident that the function in Eq. \ref{main_sf_string} satisfies all the necessary conditions to be a shape function for $r_0 = 3$ and $0<a<1$. Moreover, the function satisfies the condition $b(r_0) = r_0$. We therefore propose $b(r) = r_0 - a (r_0 - r) \left[1 + r_0~ \text{cosech}(r_0) \text{sech}(r_0) \right]$ as a new viable shape function.

\begin{figure}[H]
    \centering
    \includegraphics[width=\textwidth]{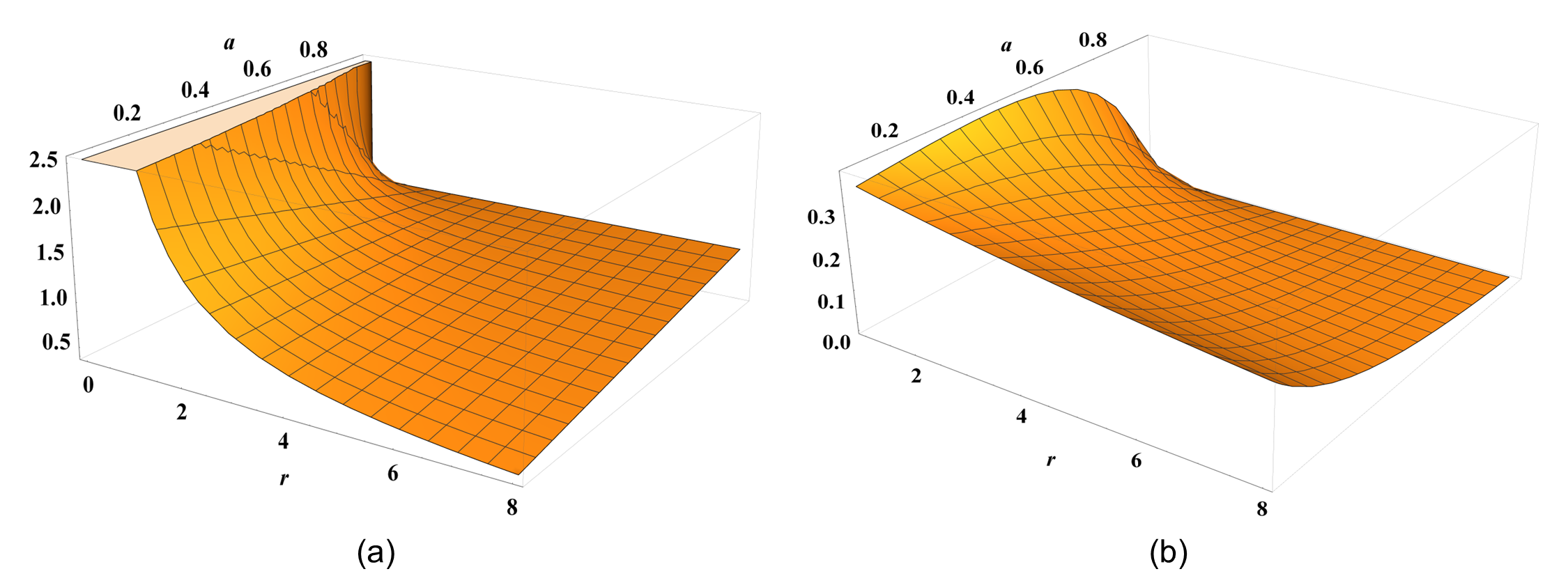}
    \caption{ \centering Profile of the (a) Asymptotic flatness~ $\frac{b(r)}{r} \rightarrow 0$ as $r \rightarrow \infty$, and (b) Flaring out condition~ $\frac{b(r) - b'(r) r}{b^2}>0$~ vs. $r$ for $0<a<1$}
    \label{ch4fig:1sf}
\end{figure}

The plots of the asymptotic flatness $b(r)/r \rightarrow 0 $ as $r \rightarrow \infty$, and flaring out condition $\frac{b(r) - b'(r) r}{b^2}>0$ are shown in Figure \ref{ch4fig:1sf}.

With the obtained viable shape function, next the traversable wormhole solutions are analysed in the framework of $f(R)$ gravity with a background of string cloud.

\section{Traversable wormholes in $f(R)$ gravity}\label{sec3}
A general form of the action in $f(R)$ modified theories is \cite{lobo2009wormhole}
    
    \begin{equation}
        S = \int d^4 x \sqrt{-g} \left[ f(R) + \mathcal{L}_m \right]
        \label{fract}
    \end{equation}
    
Using the metric formalism of $f(R)$ gravity, the modified EFEs are obtained as
    
    \begin{equation}
    FR_{\mu\nu}-\frac{1}{2}f(R)\,g_{\mu\nu}-\nabla_\mu \nabla_\nu
    F+g_{\mu\nu}\Box F=\,T^m_{\mu\nu} \,,
    \label{eq:f_R}
    \end{equation}
where $F\equiv df/dR$, and $T^m_{\mu\nu} = \frac{-2}{\sqrt{-g}} \frac{\delta \mathcal{L}_m}{\delta g^{\mu \nu}}$ is the stress-energy tensor of background matter. We consider that the spherically symmetric space-time is represented by the line-element in Eq. (\ref{mtle}). Contracting Eq. (\ref{eq:f_R}) yields:
    \begin{equation}
    FR-2f(R)+3\Box F=T
    \label{eq:f_R_box}
    \end{equation}
     
Here, $T$ is the trace of the stress-energy tensor of matter and $\Box F$ is:
    \begin{align}
        \Box F= \frac{1}{\sqrt{-g}} \partial_\mu (\sqrt{-g} g^{\mu \nu} \partial_\nu F) = \left(1-\frac{b}{r}\right)\left[F''
    -\frac{b'r-b}{2r^2(1-b/r)}\,F'+\frac{2F'}{r}\right]
    \end{align}
with $F'=d f(R)/d R$ and $b'=d\,b(r)/dr$. Substituting Eq. (\ref{eq:f_R_box}) in Eq. (\ref{eq:f_R}) yields the modified EFEs as:
    \begin{equation}
    G_{\mu\nu}\equiv R_{\mu\nu}-\frac{1}{2}g_{\mu\nu} R= T^{{\rm
    eff}}_{\mu\nu} \,
        \label{eq:f_R_fe}
    \end{equation}

Here, $T^{{\rm eff}}_{\mu\nu}$ is the effective stress-energy tensor, responsible for the energy condition violations and is generally interpreted as a gravitational fluid. $T^{{\rm eff}}_{\mu\nu}$ comprises of the matter stress energy tensor $T^m_{\mu\nu}$ and the curvature stress-energy tensor $T^{{\rm c}}_{\mu\nu}$ given by
    \begin{align}
    T^{c}_{\mu\nu}=\frac{1}{F}\left[\nabla_\mu \nabla_\nu F
    -\frac{1}{4}g_{\mu\nu}\left(RF+\Box F+T\right) \right]
        \label{eq:7}
    \end{align}

Assuming that the geometry of the wormhole is threaded by an anisotropic distribution of matter
    \begin{equation}
    T_{\mu\nu}=(\rho+p_t)U_\mu \, U_\nu+p_t\,
    g_{\mu\nu}+(p_r-p_t)\chi_\mu \chi_\nu \,,
    \end{equation}
where $U^\mu$ is the four-velocity, and $\chi^\mu$ represents a unit space-like vector.

With the line element in Eq. (\ref{mtle}), the modified EFEs can be expressed as the following \cite{lobo2009wormhole}
    
    \begin{align}
    \rho=\frac{Fb'}{r^2}
    \label{generic1}
    \end{align}
    \begin{align}
    p_r=-\frac{bF}{r^3}+\frac{F'}{2r^2}(b'r-b)-F''\left(1-\frac{b}{r}\right)
    \label{generic2}
    \end{align}
    \begin{align}
    p_t=-\frac{F'}{r}\left(1-\frac{b}{r}\right)+\frac{F}{2r^3}(b-b'r)
    \label{generic3}
    \end{align}

The Ricci scalar is given as $R=\frac{2 b'}{r^2}$. With the explicit forms of energy density $\rho$, radial pressure $p_r$, and transverse pressure $p_t$ in Eqs. \eqref{generic1}-\eqref{generic3}, the various energy conditions, stability and the amount of exotic matter required for the wormhole configuration can be analysed.

\subsection{Cloud of Strings as a source}\label{subsec3_1}
A cloud of strings is analogous to the perfect fluid models of gas and dust. However, the difference is that it comprises one-dimensional objects extended along some specific direction. The string cloud can exist in different geometrical configurations such as planar, axisymmetric, or spherical  \cite{gracca2018cloud}. A general solution for the spherical distribution of the cloud of strings was reported in \cite{gracca2018cloud}. In addition, thermodynamic properties of string gas has been reported in \cite{mertens2014relevance}. For a spherically symmetric string cloud in four-dimensions, the energy momentum tensor has the following non-null components \cite{gracca2018cloud}
\begin{equation}
\label{string_ene_mom}
{T^{t}}_{t} = {T^{r}}_{r} = - \frac{\eta^2}{r^2}
\end{equation}
where $\eta$ is a constant related to the total energy of the string cloud.
With the form of the shape function as described in Eq. \eqref{main_sf_string}, and using Eq. \eqref{generic1}, we get
\begin{equation}
\label{form_F_r}
F(r) = -\frac{\eta ^2}{a \left( r_0~ \text{cosech}(r_0)~ \text{sech}~ (r_0) + 1 \right) }
\end{equation}

With the obtained $F(r)$ and the field equations Eqs. \eqref{generic1}-\eqref{generic3}, the numerical analyses are conducted to obtain the energy conditions, check the stability, and estimate the amount of exotic matter.

For any observer traversing a time-like curve to detect the energy density of the matter field to be positive, the stress-energy tensor of matter must adhere to some sets  of inequalities known as the energy conditions in GR \cite{curiel2017primer,Baruah_2019}. The weak energy condition (WEC) implies $\rho \geq 0$, the NEC implies $\rho + p_r \geq 0$, and $\rho + p_t \geq 0$, whereas the strong energy condition (SEC) implies $\rho + p_r + 2 p_t \geq 0$.
The profile of energy conditions with a background of string cloud is presented with the throat radius fixed at $r_0=3$. Emphasis is made on the effect of the string cloud constant $\eta$ on the energy conditions. Results are discussed by fixing the parameter $a=0.5$. For the whole range of $a$, $0<a<1$, the energy conditions show similar behaviour. Previous studies indicate that the constant associated with the total energy of the string cloud should be small \cite{letelier1979clouds,richarte2008traversable}.

\begin{figure}[H]
    \centering
    \includegraphics[width=\textwidth]{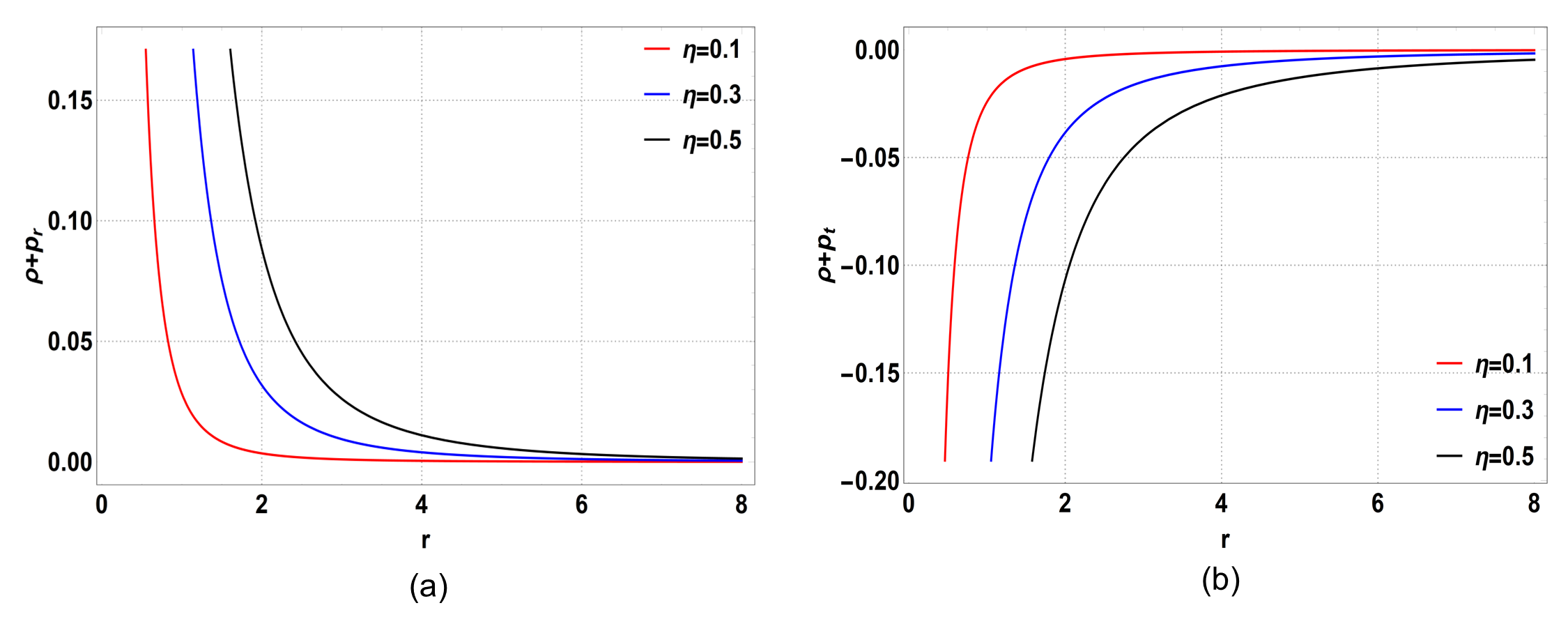}
    \caption{ \centering Profile of the NEC terms (a) $\rho + p_r$, and (b) $\rho + p_t$ respectively vs. $r$ with $b(r)$ as in Eq. \ref{main_sf_string}}
    \label{stringNEC}
\end{figure}

Figure \ref{stringNEC} shows the profile of the NEC terms $\rho + p_r$ and $\rho + p_t$. It can be observed that the first NEC term is satisfied at the throat $r_0=3$ for the considered values of the $\eta$ as $\eta = 0.1, \, 0.3, \, \text{and}~ 0.5$. However, the second NEC term  $\rho + p_t$ is violated for all the values of $\eta$. Owing to the violation of the second NEC term, as a whole the NEC is inferred to be violated.

\begin{figure}[H]
    \centering
    \includegraphics[width=\textwidth]{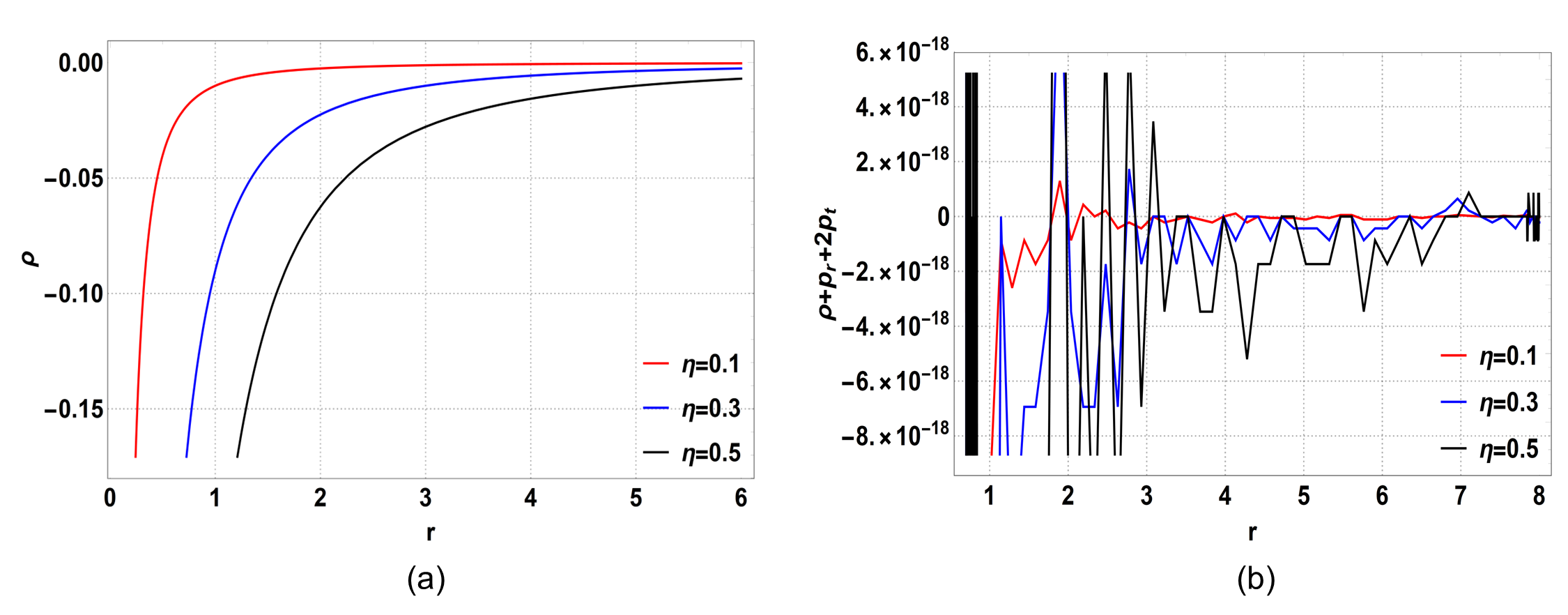}
    \caption{ \centering Profile of the (a) WEC $\rho$, and (b) SEC $\rho + p_r +2p_t$ vs. $r$ respectively with $b(r)$ as in Eq. \ref{main_sf_string}}
    \label{stringWEC_SEC}
\end{figure}

From figure \ref{stringWEC_SEC}, it can be observed that the WEC is violated at the wormhole throat for all the values of $\eta$. It is also evident from the fact that the non-null components of the stress-energy tensor of the string cloud comes with a minus sign as shown in Eq. \eqref{string_ene_mom}. Moreover, it is seen that the SEC exhibits an oscillatory (indeterminate) behaviour along the radial co-ordinate $r$. With increasing $r$, the oscillation between positive and negative values decreases; however, the behaviour extends asymptotically.

In addition to the energy conditions, analysing two other parameters, viz. the equation of state (EoS) parameter $\omega  = p_r/\rho$, and anisotropy parameter $\Delta = p_t - p_r$ turns out to be useful. Information about the nature of the matter source threading the wormhole geometry can be obtained from the EoS parameter, and the attractive or repulsive nature of the space-time geometry (geometrical viability) can be understood by the anisotropy parameter.

\begin{figure}[H]
    \centering
    \includegraphics[width=\textwidth]{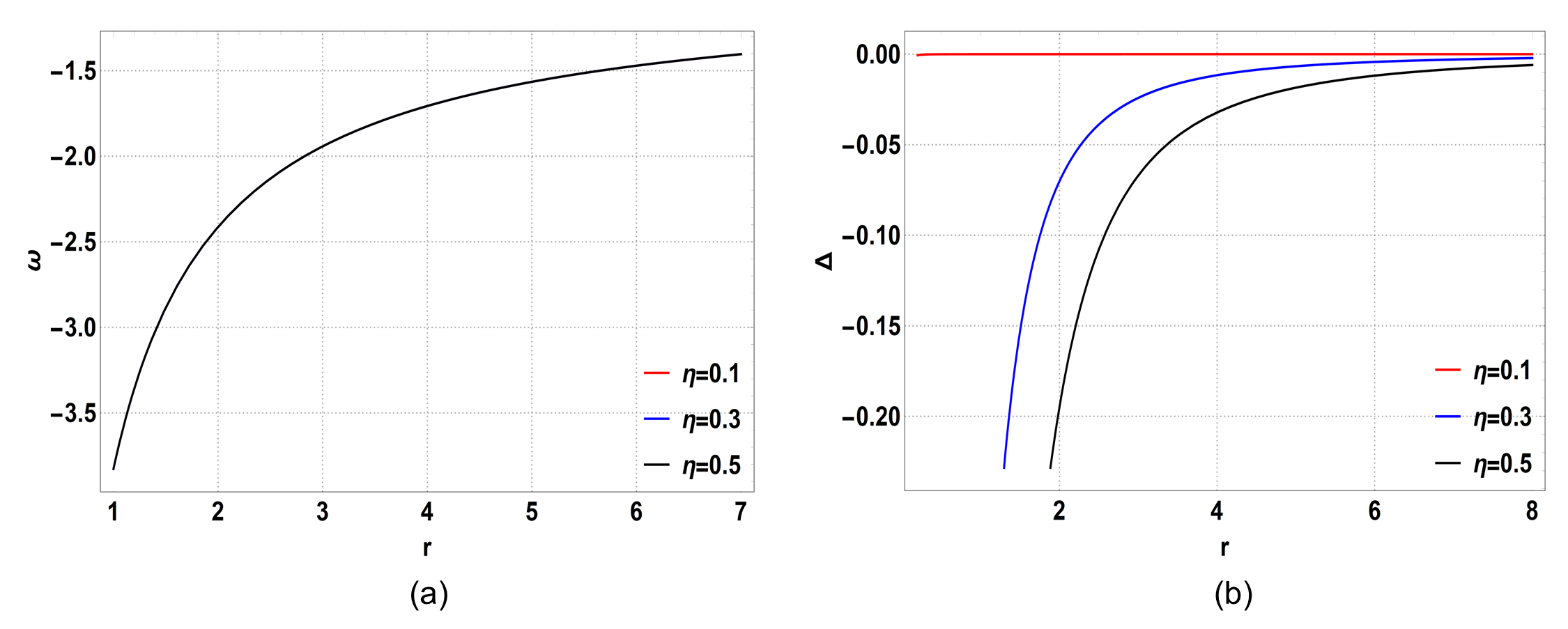}
    \caption{ \centering Profile of the (a) EoS parameter $\omega$, and (b) anisotropy parameter $\Delta$ vs. $r$ respectively with $b(r)$ as in Eq. \ref{main_sf_string}}
    \label{stringEoS_Ani}
\end{figure}

Figure \ref{stringEoS_Ani} shows the behaviour of the EoS  and the anisotropy parameters. Near the wormhole throat, the EoS parameter is $\omega < -1$, signifying a phantom-like behaviour of the string cloud for all values of $\eta$. The anisotropy parameter $\Delta < 0$ near the wormhole throat, signifying an attractive nature of the space-time geometry. The summary of the energy conditions are presented in Table \ref{tab1}. 

\begin{table}
\centering
\caption{Summary of the energy conditions discussed in Sec. \ref{subsec3_1}} \label{tab1}
\begin{tabular}{@{}lll@{}}
\hline
Terms & \makecell{Result} & Interpretation \\
\hline
$\rho$ & \makecell{$<0$ near throat, \\ for $\eta=0.1, 0.3, 0.5$} & \makecell{WEC violated \\ at throat} \\
\hline
$\rho + p_r$ & \makecell{$>0$ near throat, \\ for $\eta=0.1, 0.3, 0.5$} & \makecell{NEC satisfied \\ at throat}  \\
\hline
$\rho + p_t$ & \makecell{$<0$ near throat, \\ for $\eta=0.1, 0.3, 0.5$} & \makecell{NEC violated \\ at throat} \\
\hline
$\rho + p_r + 2 p_t$ & \makecell{oscillates, \\ for $\eta=0.1, 0.3, 0.5$} & \makecell{SEC indeterminate} \\
\hline
$\omega$ & \makecell{$< -1$ near throat, \\ for $\eta=0.1, 0.3, 0.5$} & \makecell{phantom-like source \\ at throat} \\
\hline
$\Delta$ & \makecell{$< 0$ near throat, \\ for $\eta=0.1, 0.3, 0.5$} & \makecell{attractive geometry \\ at throat} \\
\hline
\end{tabular}
\end{table}

First reported in the context of neutron stars \cite{tolman1987relativity,oppenheimer1939massive}, the Tolman-Oppenheimer-Volkov (TOV) equation provides information regarding the stability of stellar structures. In order to probe the stability of wormholes in terms of the hydrostatic, gravitational, and anisotropic forces in the space-time, a more generalized version of the formalism was developed in \cite{gorini2008tolman}. The generalized TOV equation \cite{gorini2008tolman,ponce1993limiting} is given as
    
    \begin{equation}\label{eq:tov}
    -\frac{dp_{r}}{dr}-\frac{\epsilon'(r)}{2}(\rho+p_{r})+\frac{2}{r}(p_{t}-p_{r})=0,
    \end{equation}
where $\epsilon(r)=2\Phi(r)$. $F_{\mathrm{h}}$ represents the hydrostatic force, $F_{\mathrm{g}}$ the gravitational force, and $F_{\mathrm{a}}$, the anisotropic force. These three terms of the TOV equation can determine the equilibrium anisotropic mass distribution \cite{ponce1993limiting} in that stable stellar structures satisfy Eq. \eqref{eq:tov}.
    
    \begin{equation}\label{stabcomp}
    F_{\mathrm{h}}=-\frac{dp_{r}}{dr},\;\;\;\;\;\;\;\;F_{\mathrm{a}}=\frac{2}{r}(p_{t}-p_{r}), \;\;\;\;\;\;\;\;F_{\mathrm{g}}=-\frac{\epsilon^{'}}{2}(\rho+p_{r})
    \end{equation}
    
Owing to the constant red-shift function $\Phi'(r)=0$, the gravitational force is $F_{\mathrm{g}}=0$ in the analysis.
 
Using the averaged null energy condition, $\int_{\lambda_1}^{\lambda_2} T_{ij}k^ik^j d \lambda \geq 0$, evaluated along the radial coordinate $r$, the amount of exotic matter in wormhole space-times can be estimated. However, the amount of energy condition violating matter can be estimated in a more generalised manner by using a volume integral instead of the line integral, namely the volume integral quantifier (VIQ) \cite{visser2003traversable,kar2004quantifying,lobo2013new}, which is defined as
    \begin{equation}
    \label{VIQ}
    I_v = \oint [\rho+p_r] dV = 8\pi\int_{r_0}^{s}(\rho+p_r)r^2 dr
    \end{equation}
    
After matching the wormhole space-time with an exterior metric and cutting off the stress-energy tensor at some $r=s$ away from the throat, an estimate of the amount of NEC violating matter can be obtained by using the VIQ. For arbitrarily small quantities of NEC violating matter, the requirement is $I_v \rightarrow 0$ as $s \rightarrow r_0$ \cite{visser2003traversable,kar2004quantifying}.

\begin{figure}[!ht]
\centering
\includegraphics[width=\textwidth]{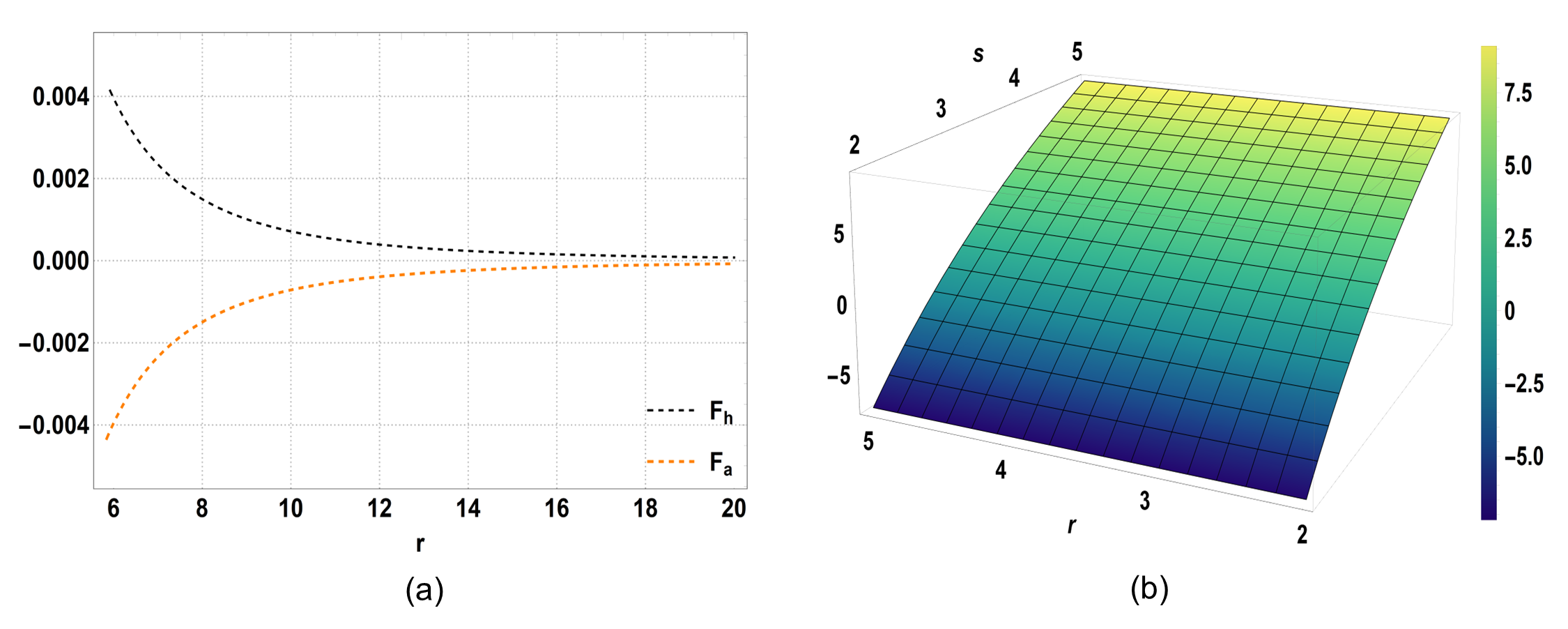}
\caption{\centering Profile of the (a) $F_{\mathrm{h}}$, and $F_{\mathrm{a}}$ vs. $r$ (b) VIQ with $b(r)$ as in Eq. \ref{main_sf_string}}
\label{stringTOV_VIQ}
\end{figure}

Figure \ref{stringTOV_VIQ} shows the terms of the TOV equation and the profile of the VIQ. It is seen that that the hydrostatic force $F_{\mathrm{h}}$, and the anisotropic force $F_{\mathrm{a}}$ cancel each other asymptotically, signifying a stable configuration. In addition, it can be observed that the VIQ, $I_v \rightarrow 0$ as $s \rightarrow r_0$, indicates that the wormhole solution is feasible with arbitrarily small amounts of exotic matter. The terms of the TOV and VIQ equations are shown with $\eta =0.5$. For $\eta= 0.1,~ \text{and}~ 0.3$ the corresponding terms of the TOV and VIQ equations depict a similar behaviour.

Thus, wormhole solutions can be formulated in a model independent manner using the novel shape function, Eq. \eqref{main_sf_string}, in the framework of $f(R)$ gravity with a string cloud background. The next motive remains to check whether a feasible form of the shape function can be obtained in a particular form of $f(R)$ gravity model with the string cloud cloud background.

\subsection{Traversable wormholes in $f(R) = \alpha R^m - \beta R^n$ gravity with a string cloud background}\label{subsec3_2}
In this section, we obtain a form of the shape function $b(r)$ by fixing $f(R)$ and analyse its viability. For the analysis, we consider a simple form of $f(R)$ gravity model as \cite{nojiri2008modified} 
\begin{equation}
\label{f_R}
f(R) = \alpha R^m - \beta R^n
\end{equation}
where $\alpha$ and $\beta$ are positive constants and $m$ and $n$ are positive integers satisfying the condition $m>n$ \cite{nojiri2008modified}. Using Eq. \eqref{string_ene_mom} in Eq. \eqref{generic1}, we obtain
\begin{equation}
\label{obtain_sf_1}
-\frac{\eta^2}{r^2} = \frac{F b'}{r^2}
\end{equation}

Considering the $f(R)$ model as described in Eq. \eqref{f_R}, and obtaining $F$, the shape function can be found from Eq. \eqref{obtain_sf_1}. For simplifying the calculations, the integers $m$ and $n$ are set as $m=2$ and $n=1$ satisfying the condition $m>n$. Again, using the expression for the Ricci scalar $R=\frac{2b'}{r^2}$, $F$ is found in terms of $r$. Eq. \eqref{obtain_sf_1} reduces to a quadratic equation in $b'$ of the form
\begin{equation}
\label{obtain_sf_2}
4 \alpha {b'}^2 - b' \beta r^2 + \eta^2 r^2 = 0
\end{equation}

Solving Eq. \eqref{obtain_sf_2} and considering only the positive root for mathematical feasibility, we get
\begin{equation}
\label{obtain_sf_3}
b' = \frac{\beta  r^2+\sqrt{\beta ^2 r^4-16 \alpha  \eta ^2 r^2}}{8 \alpha }
\end{equation}

Integrating Eq. \eqref{obtain_sf_3}, we get
\begin{equation}
\label{obtain_sf_4}
b = \frac{1}{8 \alpha} \left[ \frac{\beta  r^3}{3}+\frac{\left(\beta ^2 r^2-16 \alpha  \eta ^2\right) \sqrt{\beta ^2 r^4-16 \alpha  \eta ^2 r^2}}{3 \beta ^2 r} \right] + c
\end{equation}
where $c$ is a constant of integration. In order to evaluate the constant $c$, we use the condition at the wormhole throat, $b(r_0) = r_0$. This leads to the following form of the shape function 
\begin{align}
\label{obtain_sf_final}
b&=  r_0 + \frac{1}{24 \alpha  \beta ^2} \left[ \frac{\beta ^3 r^4+\left(\beta ^2 r^2-16 \alpha  \eta ^2\right) \sqrt{\beta ^2 r^4-16 \alpha  \eta ^2 r^2}}{r} \right. \nonumber \\
&- \left. \frac{\beta ^3 r_0^4+\left(\beta ^2 r_0^2-16 \alpha  \eta ^2\right) \sqrt{\beta ^2 r_0^4-16 \alpha  \eta ^2 r_0^2}}{r_0} \right]
\end{align}

\begin{figure}[!ht]
\centering
\includegraphics[width=\textwidth]{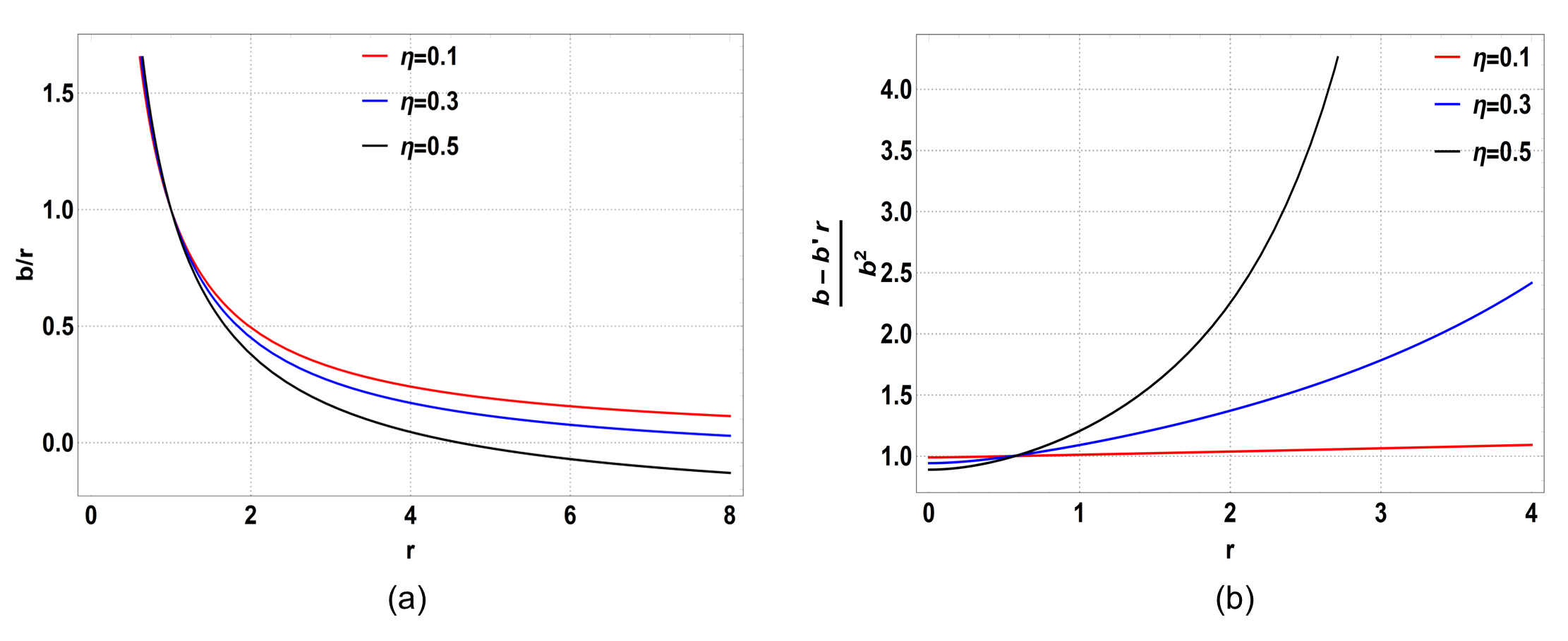}
\caption{\centering Profile of the (a) asymptotic flatness $\frac{b}{r}$ and (b) flaring out condition $\frac{b-b' r}{b^2}$ vs. $r$ with $r_0 = 1$, $\alpha = -0.5$ and $\beta = -0.8$, for $b(r)$ in Eq. \ref{obtain_sf_final}}\label{Sfstrng}
\end{figure}

With the shape function in Eq. \eqref{obtain_sf_final}, the various conditions required for the viability of the shape function as described in Sec \ref{sec2} are analysed. It is interesting to note that the asymptotic flatness $\frac{b}{r} \rightarrow 0$ as $r \rightarrow \infty$, and the flaring out condition $\frac{b-b' r}{b^2} > 0$ are satisfied only for negative values of the constants $\alpha$ and $\beta$ ($\alpha =-0.5$ and $\beta =-0.8$). The profile of the asymptotic flatness and flaring out condition are shown in Figure \ref{Sfstrng} with different values of $\eta$.
With the shape function in Eq. \eqref{obtain_sf_final}, and the $f(R)$ model described by Eq. \eqref{f_R}, the energy conditions, stability and the amount of exotic matter required for the wormhole configuration are analysed next.

Using Eqs. \eqref{generic1}-\eqref{generic3}, the various energy conditions are analysed with the throat at $r_0 = 1$, and by fixing the model parameters of the $f(R)$ gravity model (Eq. \eqref{f_R}) as, $m=2, \, n=1$, $\alpha =-0.5$ and $\beta =-0.8$.
\begin{figure}[!ht]
\centering
\includegraphics[width=\textwidth]{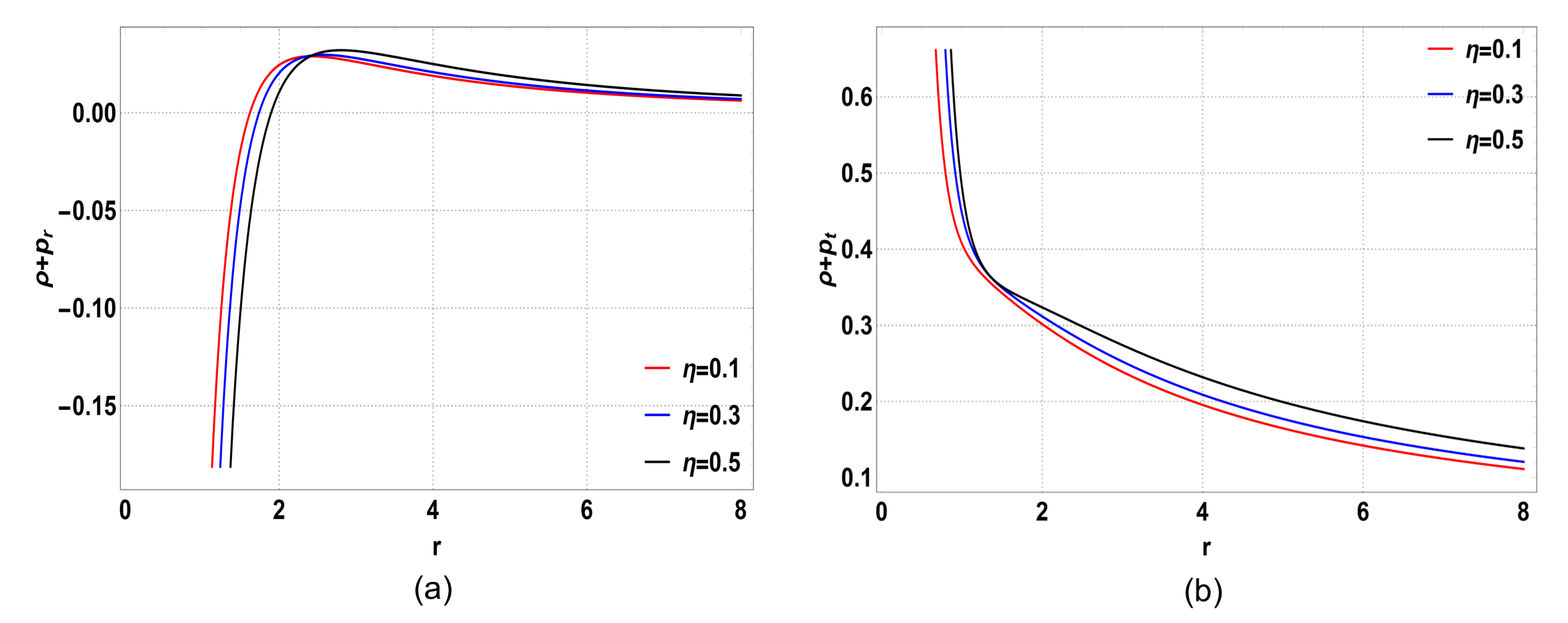}
\caption{\centering Profile of the NEC terms (a) $\rho + p_r$ and (b) $\rho + p_t$ vs. $r$ with $b(r)$ as in Eq. \ref{obtain_sf_final}}\label{second_NEC}
\end{figure}

\begin{figure}[!ht]
\centering
\includegraphics[width=\textwidth]{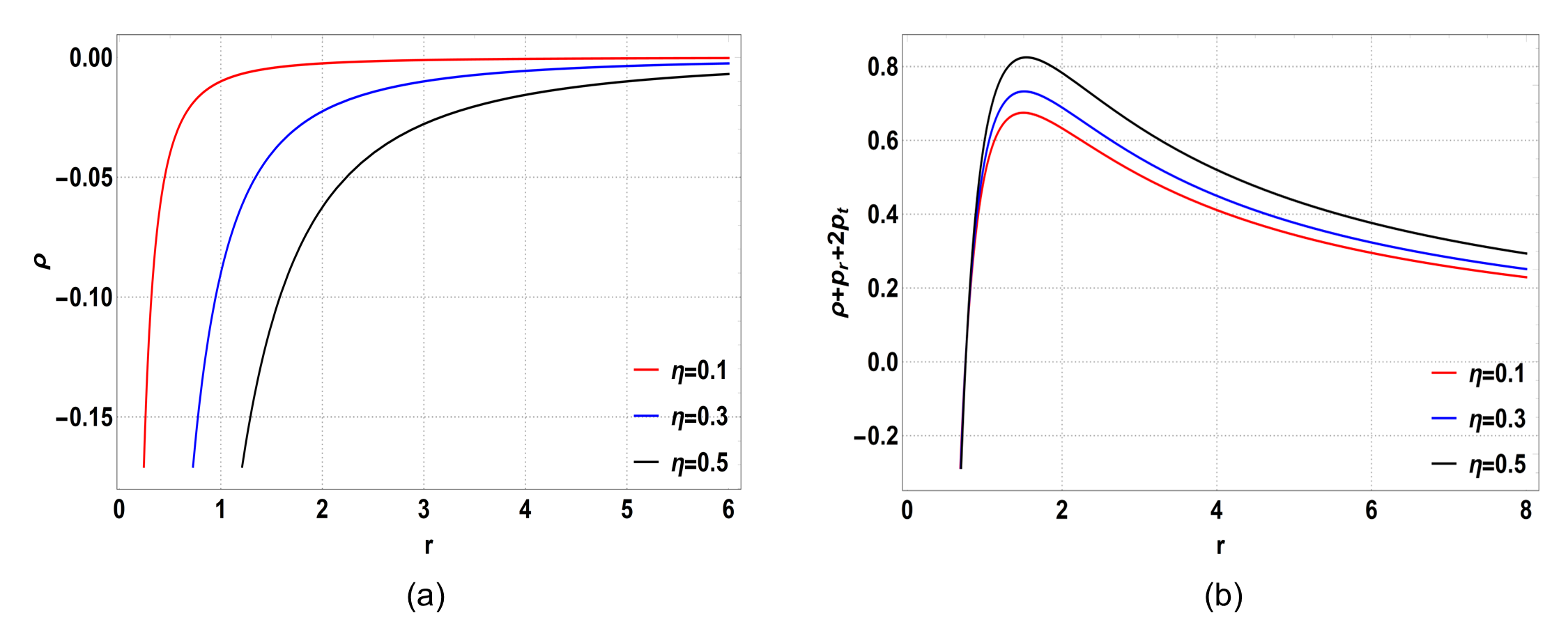}
\caption{\centering Profile of the (a) WEC $\rho$ and (b) SEC $\rho + p_r +2p_t$ vs. $r$ with $b(r)$ as in Eq. \ref{obtain_sf_final}}\label{second_WEC_SEC}
\end{figure}

\begin{figure}[!ht]
\centering
\includegraphics[width=\textwidth]{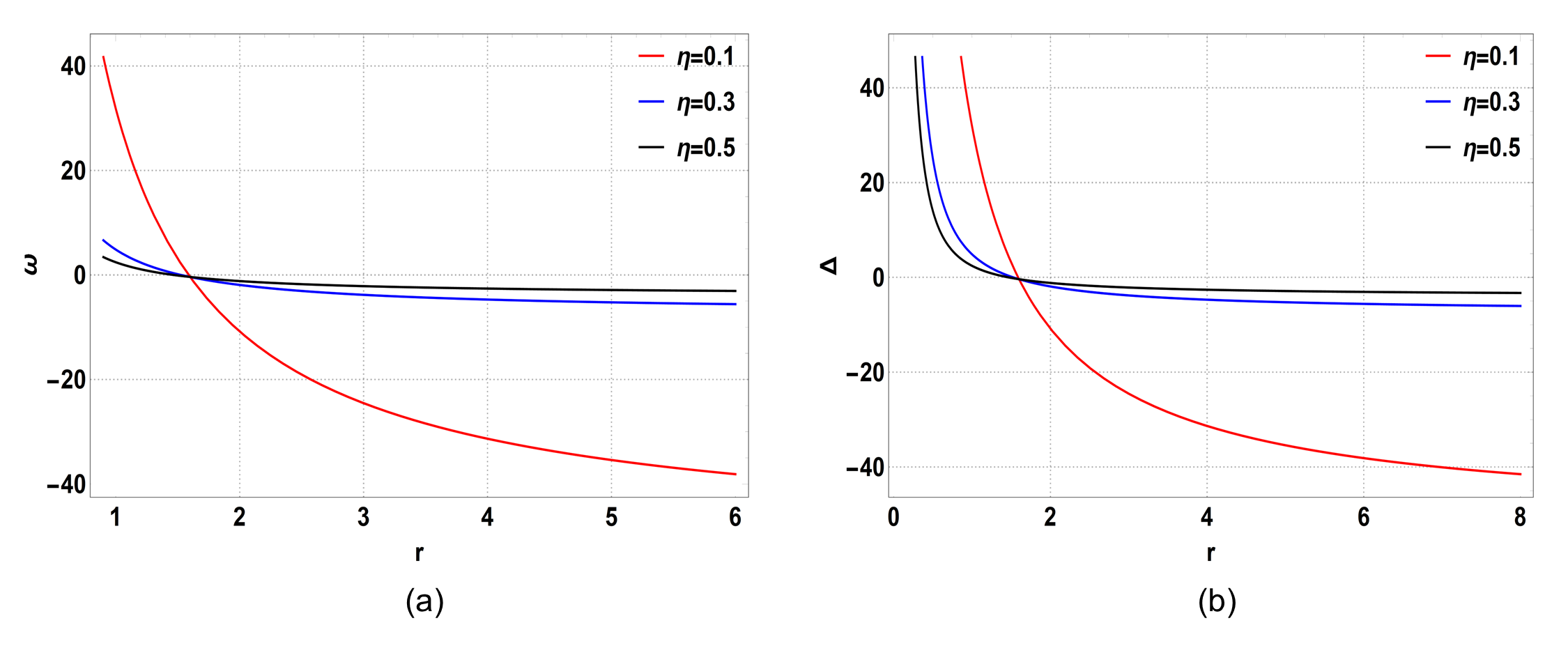}
\caption{\centering Profile of the (a) EoS parameter $\omega$ and (b) anisotropy parameter $\Delta$ vs. $r$ with $b(r)$ as in Eq. \ref{obtain_sf_final}}\label{second_EoS_Aniso}
\end{figure}

Figure \ref{second_NEC} shows the NEC terms $\rho + p_r$, and  $\rho + p_t$. It can be observed that the first NEC term $\rho + p_r$ is violated at the wormhole throat. The second NEC term $\rho + p_t$ is satisfied at the wormhole throat. However, owing to violation to the first NEC term, as a whole the NEC is inferred to be violated.

Figure \ref{second_WEC_SEC} shows the WEC and the SEC. It can be seen that the WEC is violated at the wormhole throat, and this violation can again be attributed to the negative sign associated with the non-null components of the stress-energy tensor of the string cloud as in Eq. \eqref{string_ene_mom}. The SEC is satisfied at the throat and also asymptotically, which is again an interesting point to note as in $f(R)$ gravity, the SEC should be asymptotically violated to account for the late-time accelerated expansion of the universe.

\begin{table}[!ht]
\centering
\caption{Summary of the energy conditions discussed in Sec. \ref{subsec3_2}}\label{tab2}
\begin{tabular}{@{}lll@{}}
\hline
Terms & \makecell{Result} & \makecell{Interpretation} \\
\hline
$\rho$ & \makecell{$<0$ near throat, \\ for $\eta=0.1, 0.3, 0.5$} & \makecell{WEC violated \\ at throat} \\
\hline
$\rho + p_r$ & \makecell{$<0$ near throat, \\ for $\eta=0.1, 0.3, 0.5$} & \makecell{NEC violated \\ at throat}  \\
\hline
$\rho + p_t$ & \makecell{$>0$ near throat, \\ for $\eta=0.1, 0.3, 0.5$} & \makecell{NEC satisfied \\ at throat} \\
\hline
$\rho + p_r + 2 p_t$ & \makecell{$>0$ near throat, \\ for $\eta=0.1, 0.3, 0.5$} & \makecell{SEC satisfied \\ at throat} \\
\hline
$\omega$ & \makecell{$> 0$ near throat, \\ for $\eta=0.1, 0.3, 0.5$} & \makecell{normal matter like source \\ at throat} \\
\hline
$\Delta$ & \makecell{$> 0$ near throat, \\ for $\eta=0.1, 0.3, 0.5$} & \makecell{repulsive geometry \\ at throat} \\
\hline
\end{tabular}
\end{table}

\begin{figure}[!ht]
\centering
\includegraphics[width=\textwidth]{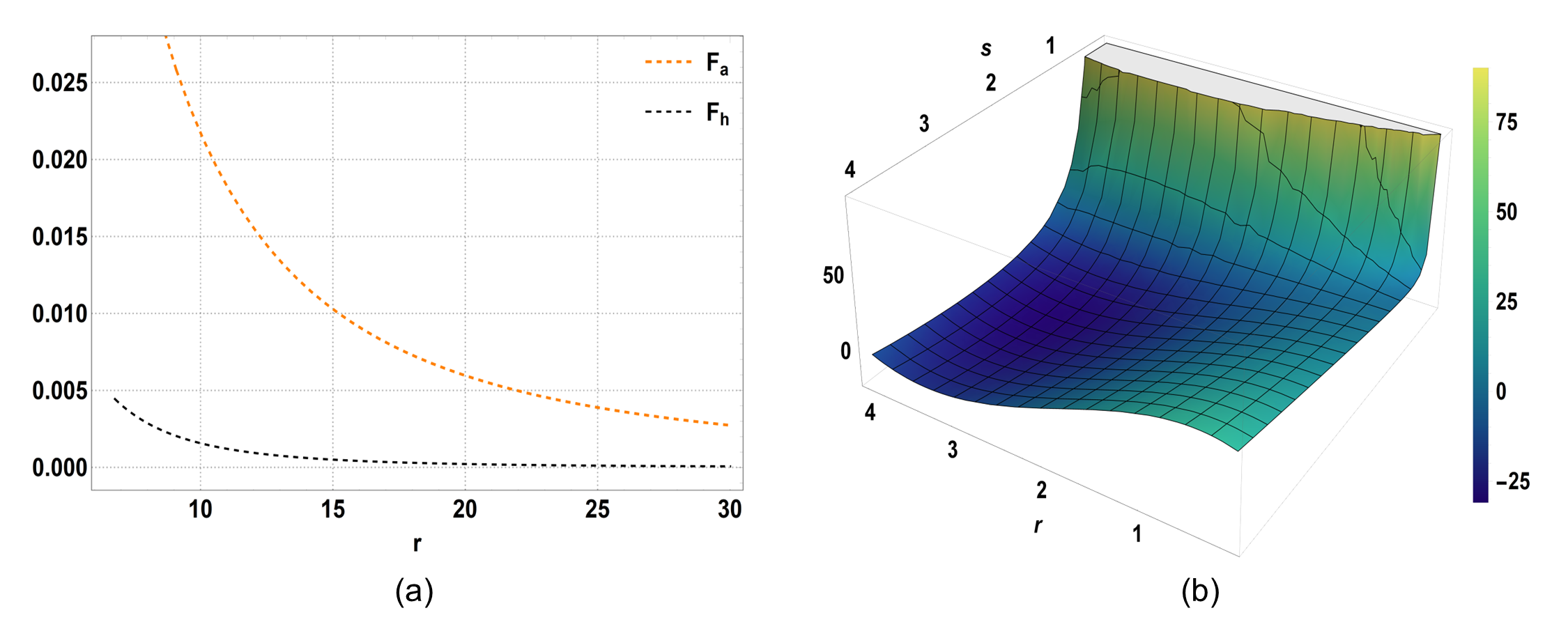}
\caption{\centering Profile of the (a) $F_{\mathrm{h}}$, and $F_{\mathrm{a}}$ vs. $r$ and (b) VIQ with $b(r)$ as in Eq. \ref{obtain_sf_final}}\label{second_TOV_VIQ}
\end{figure}

Figure \ref{second_EoS_Aniso} shows the variation of the EoS parameter $\omega$ and the anisotropy parameter $\Delta$. It can be seen that the EoS parameter $\omega$ is $\omega>0$ near the wormhole throat for all values of $\eta$, signifying that the string cloud as the source behaves like ordinary matter without any phantom-like behaviour. The anisotropy parameter $\Delta$ is positive near the wormhole throat for all the values of $\eta$, signifying a repulsive geometry at the throat. The summary of the energy conditions are presented in Table \ref{tab2}. 

Figure \ref{second_TOV_VIQ} shows the terms of the TOV equation and the VIQ. It is seen that the hydrostatic force $F_{\mathrm{h}}$, and anisotropic force $F_{\mathrm{a}}$ does not cancel each other out, signifying that the wormhole configuration is unstable. From the VIQ, it is evident that $I_v \rightarrow 0$ as $s \rightarrow r_0$, indicating that the wormhole configuration can be obtained with arbitrarily small amounts of exotic matter.

With the results presented in Sec \ref{subsec3_1} and Sec \ref{subsec3_2}, it is clear that wormhole solutions in $f(R)$ gravity with background a string cloud is feasible with characteristic violation of the NEC. However, with $f(R) = \alpha R^m - \beta R^n$ gravity, the wormhole configuration is not stable. In order to have a better understanding, the next section presents results of wormhole solution in  $f(R) = \alpha R^m - \beta R^n$ gravity supported by ordinary matter and with the form of the shape function as in Eq. \eqref{main_sf_string}. 

\subsection{Traversable wormholes in $f(R) = \alpha R^m - \beta R^n$ gravity with ordinary matter}\label{subsec3_3}
With the shape function in Eq. \eqref{main_sf_string}, the field equations Eqs. \eqref{generic1}-\eqref{generic3} are analysed with the form of the $f(R)$ model as in Eq. \eqref{f_R}. We assume that the wormhole geometry is threaded by ordinary matter with the stress-energy tensor $T^{\mu}_{\nu} = \text{daig}[-\rho(r),p_r(r),p_t(r),p_t(r)]$. Owing to the properties of the shape function, the throat radius is fixed at $r_0=3$, and the free parameter $a$ is considered for the whole range $0<a<1$. The model parameters of the $f(R) = \alpha R^m - \beta R^n$ gravity are fixed as, $\alpha=0.8$, $\beta=0.5$, and $m=2, \, n=1$. The various energy conditions, stability of the wormhole space-time and the VIQ are presented below.

Figure \ref{Third_NEC} shows the profile of the NEC terms $\rho+p_r$ and $\rho+p_t$. It is evident that that the first NEC term $\rho+p_r$ is violated at the wormhole throat. However, the second NEC term $\rho+p_t$ is satisfied at the throat. Owing to the violation of the first NEC term, as a whole the NEC is considered to be violated.

\begin{figure}[!ht]
\centering
\includegraphics[width=\textwidth]{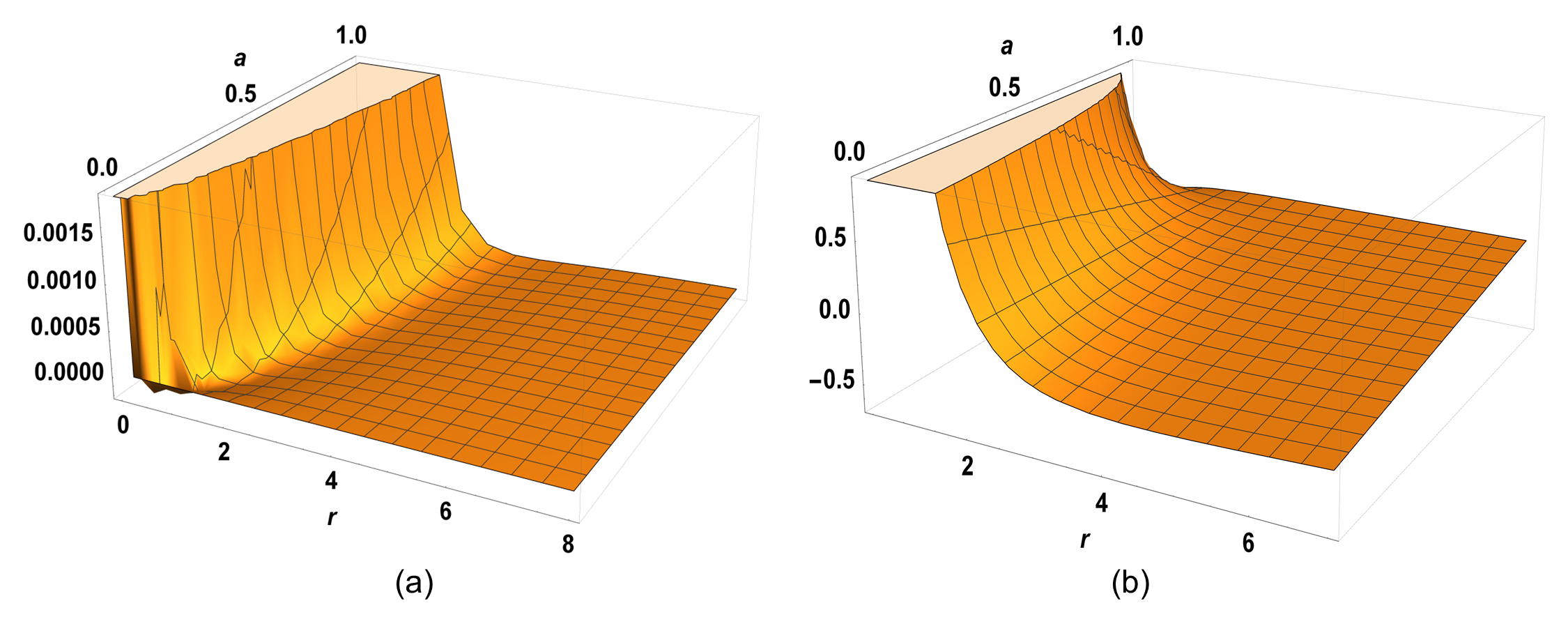}
\caption{\centering Profile of the (a) WEC $\rho$ and (b) SEC $\rho+p_r+2p_t$ vs. $r$ with $b(r)$ as in Eq. \ref{main_sf_string}}\label{Third_WEC_SEC}
\end{figure}

Figure \ref{Third_WEC_SEC} depicts the WEC and the SEC. It is seen that the WEC is violated near the wormhole throat. The SEC is satisfied at the wormhole throat and asymptotically for the whole range of $a$, $0<a<1$. It is interesting to note that even with ordinary matter as the source, the SEC is satisfied asymptotically for $f(R)=\alpha R^m - \beta R^n$ gravity.

\begin{figure}[!ht]
\centering
\includegraphics[width=\textwidth]{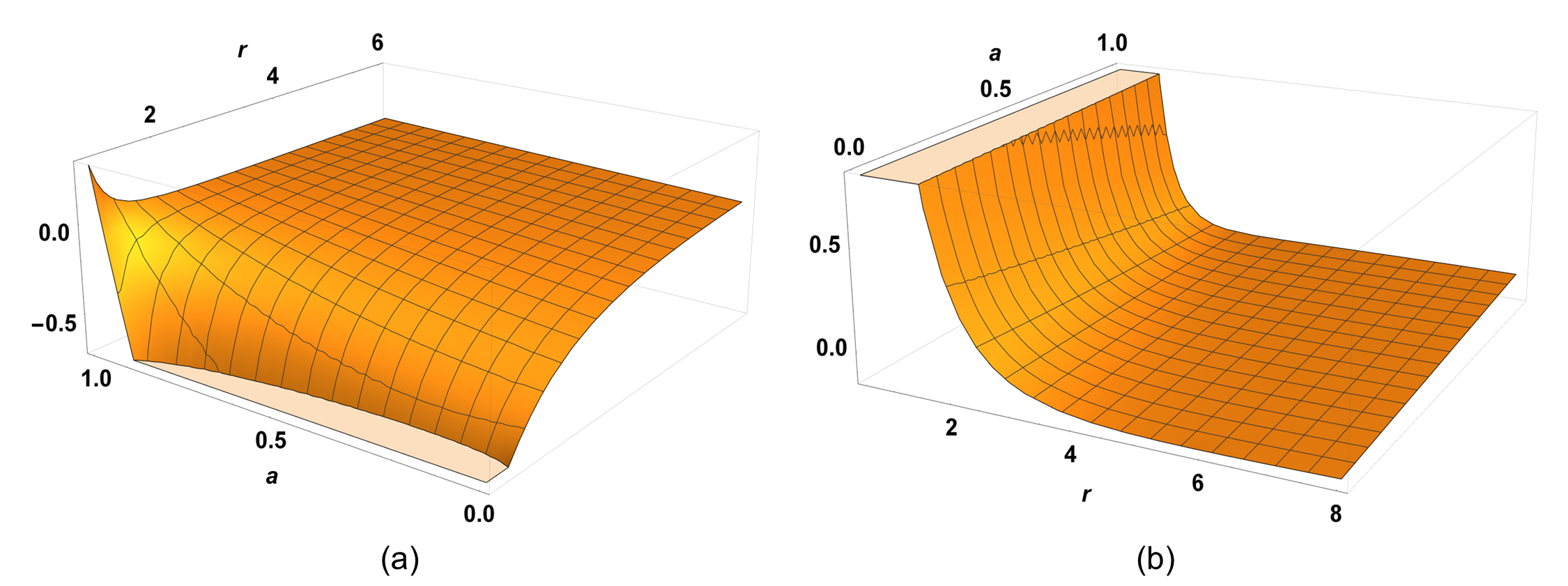}
\caption{\centering Profile of the NEC terms (a) $\rho+p_r$ and (b) $\rho+p_t$ vs. $r$ with $b(r)$ as in Eq. \ref{main_sf_string}}\label{Third_NEC}
\end{figure}

\begin{figure}[!ht]
\centering
\includegraphics[width=\textwidth]{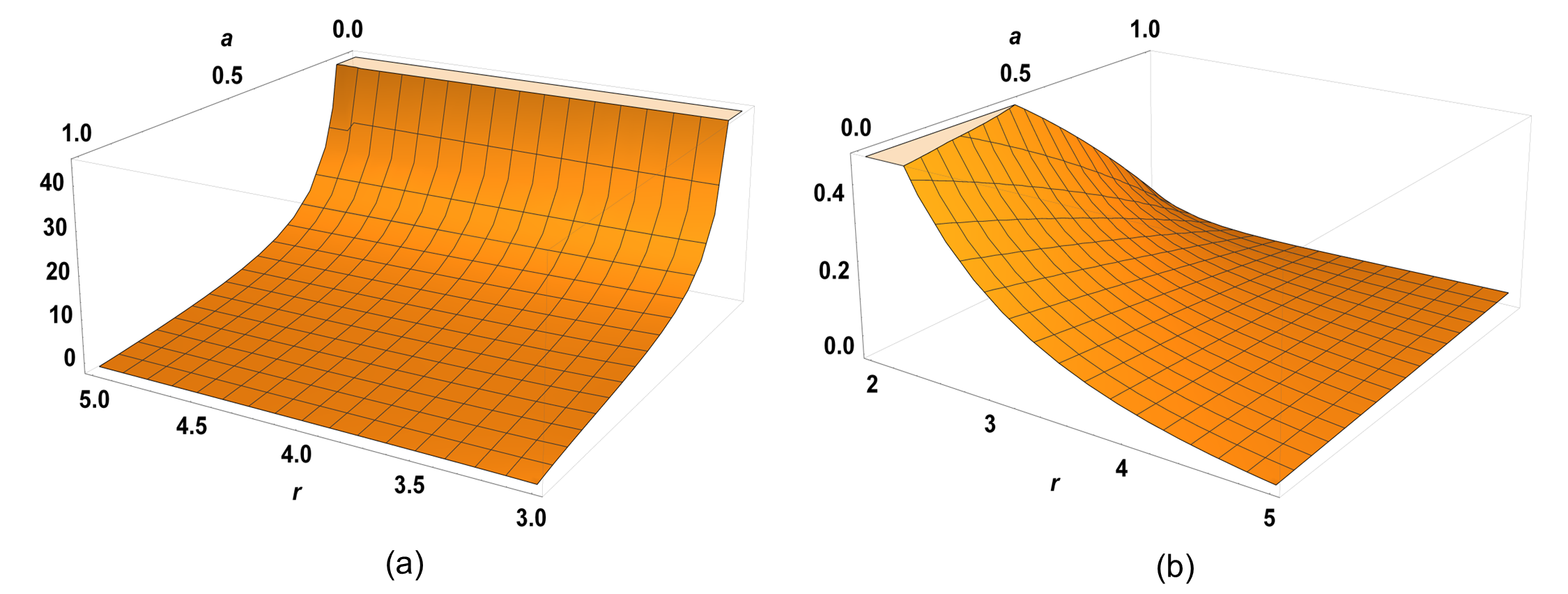}
\caption{\centering Profile of the (a) EoS parameter $\omega$ and (b) anisotropy parameter $\Delta$ vs. $r$ with $b(r)$ as in Eq. \ref{main_sf_string}}\label{Third_EoS_Aniso}
\end{figure}

Figure \ref{Third_EoS_Aniso} shows the variation of the EoS parameter and the anisotropy parameter. It can be observed that as expected, the EoS parameter is $\omega>0$ near the wormhole throat signifying that the source threading the wormhole geometry has no phantom-like behaviour. The anisotropy parameter $\Delta > 0$ near the throat, signifying a repulsive geometry at the throat. The energy conditions are summarised in Table \ref{tab3}. \\

\begin{figure}[!ht]
\centering
\includegraphics[width=\textwidth]{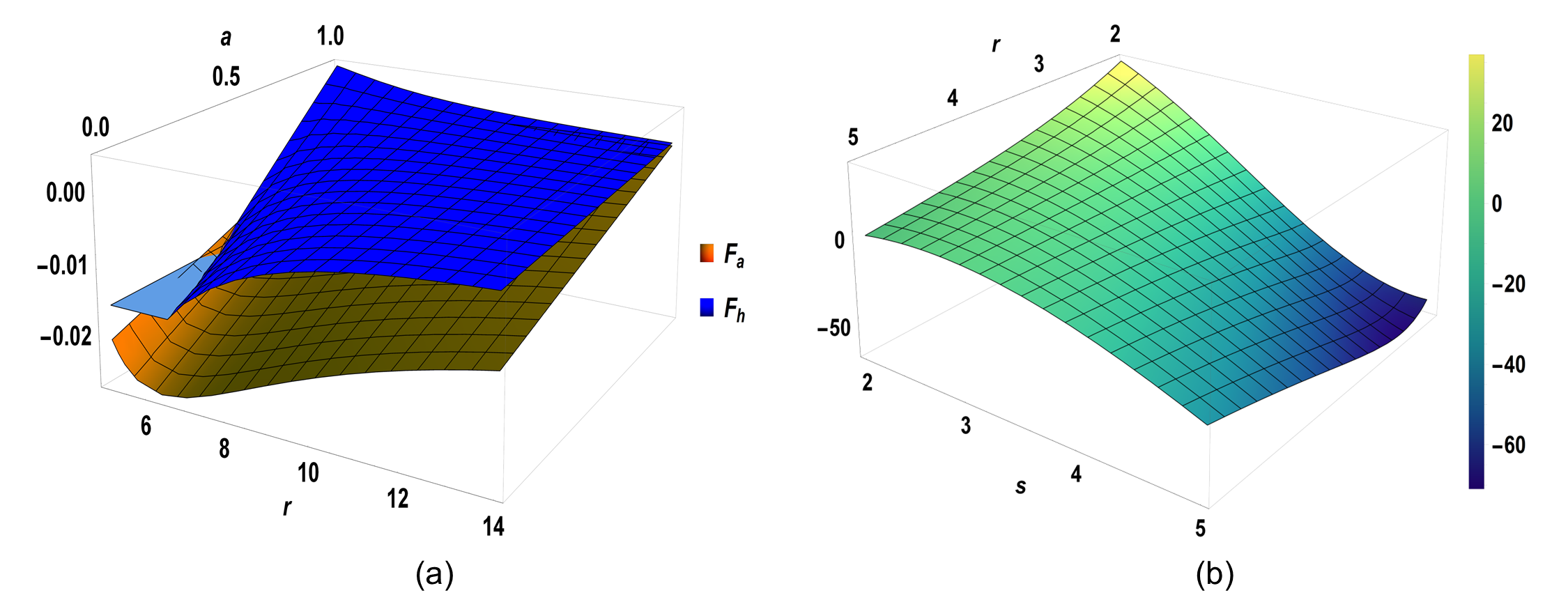}
\caption{\centering Profile of the (a) $F_{\mathrm{h}}$, and $F_{\mathrm{a}}$ vs. $r$ and (b) VIQ with $b(r)$ as in Eq. \ref{main_sf_string}}\label{Third_TOV_VIQ}
\end{figure}

Figure \ref{Third_TOV_VIQ} shows the corresponding terms of the TOV equation, and the VIQ. It is seen that both the hydrostatic force $F_{\mathrm{h}}$, and the anisotropic force $F_{\mathrm{a}}$ does not cancel each other out asymptotically for the whole range of $0<a<1$, signifying that the wormhole configuration is not stable. From the VIQ it is evident that $I_v \rightarrow 0$ as $s \rightarrow r_0$, indicating that the wormhole solution can be obtained with arbitrarily small amount of exotic matter. However, the VIQ is only shown for $a=0.5$, and for the whole range of $0<a<1$, the VIQ has a similar profile. 


\section{Discussion and Conclusion}\label{sec4}
In this study, we investigated the traversable wormholes in the framework of $f(R)$ gravity with a background of string clouds. A novel form of the shape function using Pad\'{e} approximation was proposed for the analysis. Using this shape function a specific form of $F(r)$ was obtained in Eq. \eqref{form_F_r} in order to analyse the EFEs. However, it is observed that the $F(r)$ depends on the parameters of the metric functions and on the string cloud parameter $\eta$. Eliminating these dependence and obtaining a cosmologically viable form of $f(R)$ remains an open issue. The results demonstrate that stable wormhole solutions with characteristic violation of the energy conditions (especially the NEC) are feasible in the framework of $f(R)$ gravity with the new shape function in the background of a string cloud. In addition, the EoS parameter near the wormhole throat indicates that the string cloud has phantom-like properties. The anisotropy parameter $\Delta < 0$ near the throat signifying an attractive geometry. In addition, the SEC shows an indeterminate behaviour as it oscillates between positive and negative values. Therefore, to have a better understanding of the wormhole solutions, traversable wormholes are considered in a string cloud background with a simple form of $f(R)$ gravity model, $f(R) = \alpha R^m - \beta R^n$. It is interesting to note that the shape function in this case can yield viable wormhole geometries for negative values of $\alpha$ and $\beta$. The results demonstrate characteristic violations of the NEC. Interestingly, the SEC is satisfied both at the throat and asymptotically. The EoS parameter $\omega > 0$ at the throat and indicates that that the string cloud behaves as normal matter. The anisotropy parameter $\Delta > 0$ near the wormhole throat and signifies a repulsive geometry at the throat. However, the TOV equation shows that the wormhole space-time is unstable. For a comparative analysis, wormholes threaded by normal matter are analysed in $f(R) = \alpha R^m - \beta R^n$ gravity, using the Pad\'{e} approximate shape function given in Eq. \eqref{main_sf_string}. The constants $\alpha$ and $\beta$ are considered with positive values. The results demonstrate characteristic violation of the NEC and shows that the wormhole space-time is unstable. The SEC is satisfied at the wormhole throat and asymptotically.

Although wormholes have not been detected yet, studying these exact solutions of the EFEs provide with far reaching insight into the nature of space-time and fundamental building blocks of the universe. Wormholes also play a crucial role in several important issues such as the cosmic censorship conjecture \cite{penrose2002golden}. Studies have reported strong indications for their presence in the form of black hole mimickers \cite{izmailov2019can,nandi2017ring}. Further studies regarding the observational constraints of these wormhole solutions remain an open issue to be addressed in the near future.



\begin{thebibliography}{99}

\bibitem{1916AbhKP1916..189S}
K.~{Schwarzschild}, Preuss. Akad. Wiss. Berlin (Math. Phys.) \textbf{1916},
  189--196 (1916)

\bibitem{Flamm:1916}
L.~Flamm, Phys. Z. \textbf{17}, 448 (1916)

\bibitem{einstein1935particle}
A.~Einstein, N.~Rosen, Phys. Rev. \textbf{48}(1), 73 (1935)

\bibitem{ellis1973ether}
H.G. Ellis, J. Math. Phys. \textbf{14}(1), 104--118 (1973)

\bibitem{bronnikov1973scalar}
K.A. Bronnikov, Acta Phys. Pol p.~B4 (1973)

\bibitem{Morris:1988cz}
M.S. Morris, K.S. Thorne, Am. J. Phys. \textbf{56}, 395--412 (1988)

\bibitem{Visser:1995cc}
M.~Visser, \emph{Lorentzian wormholes: From Einstein to Hawking} (American
  Institute of Physics Melville, NY, USA, 1996)

\bibitem{epr}
L.~Susskind, Fortschritte der Phys. \textbf{64}(1), 72--83 (2016)

\bibitem{perlmutter1999constraining}
S.~Perlmutter, M.S. Turner, M.~White, Phys. Rev. Lett. \textbf{83}(4), 670
  (1999)

\bibitem{riess2001farthest}
A.G. Riess, P.E. Nugent, R.L. Gilliland, B.P. Schmidt, J.~Tonry, M.~Dickinson,
  R.I. Thompson, T.~Budavari, S.~Casertano, A.S. Evans, et~al., Astrophys. J.
  \textbf{560}(1), 49 (2001)

\bibitem{linde1990particle}
A.~Linde, \emph{Particle physics and inflationary cosmology}, vol.~5 (CRC
  press, London, 1990)

\bibitem{starobinsky1980new}
A.A. Starobinsky, Phys. Lett. B \textbf{91}(1), 99--102 (1980)

\bibitem{copeland2006dynamics}
E.J. Copeland, M.~Sami, S.~Tsujikawa, Int. J. Mod. Phys. D \textbf{15}(11),
  1753--1935 (2006)

\bibitem{capozziello2011extended}
S.~Capozziello, M.~De~Laurentis, Phys. Rep. \textbf{509}(4-5), 167--321 (2011)

\bibitem{sotiriou2010f}
T.P. Sotiriou, V.~Faraoni, Rev. Mod. Phys. \textbf{82}(1), 451 (2010)

\bibitem{guo2014solar}
J.Q. Guo, Solar system tests of {$f(R)$} gravity.
\newblock Int. J. Mod. Phys. D \textbf{23}(04), 1450,036 (2014)

\bibitem{capozziello2015generalized}
S.~Capozziello, F.S. Lobo, J.P. Mimoso, Phys. Rev. D \textbf{91}(12), 124,019
  (2015)

\bibitem{Baruah_2019}
A.~Baruah, A.~Deshamukhya, J. Phys. Conf. Ser. \textbf{1330}(1), 012,001 (2019)

\bibitem{lobo2009wormhole}
F.S. Lobo, M.A. Oliveira, Phys. Rev. D \textbf{80}(10), 104,012 (2009)

\bibitem{furey2004wormhole}
N.~Furey, A.~DeBenedictis, Class. Quantum Gravity \textbf{22}(2), 313 (2004)

\bibitem{godani2019non}
N.~Godani, G.C. Samanta, Mod. Phys. Lett. A \textbf{34}(28), 1950,226 (2019)

\bibitem{godani2019traversable}
N.~Godani, G.C. Samanta, Int. J. Mod. Phys. D \textbf{28}(02), 1950,039 (2019)

\bibitem{azizi2013wormhole}
T.~Azizi, Int. J. Theor. Phys. \textbf{52}, 3486--3493 (2013)

\bibitem{mishra2020traversable}
A.K. Mishra, U.K. Sharma, V.C. Dubey, A.~Pradhan, Astrophys. Space Sci.
  \textbf{365}(2), 34 (2020)

\bibitem{boehmer2012wormhole}
C.G. Boehmer, T.~Harko, F.S. Lobo, Phys. Rev. D \textbf{85}(4), 044,033 (2012)

\bibitem{mustafa2022traversable}
G.~Mustafa, Z.~Hassan, P.~Sahoo, Ann. Phys. \textbf{437}, 168,751 (2022)

\bibitem{baruah2022new}
A.~Baruah, P.~Goswami, A.~Deshamukhya, Int. J. Mod. Phys. D p. 2250119 (2022)

\bibitem{baruah2023non}
A.~Baruah, P.~Goswami, A.~Deshamukhya, New Astron. \textbf{99}, 101,956 (2023)

\bibitem{mukhi2011string}
S.~Mukhi, Class. Quantum Gravity \textbf{28}(15), 153,001 (2011)

\bibitem{kalb1974classical}
M.~Kalb, P.~Ramond, Phys. Rev. D \textbf{9}(8), 2273 (1974)

\bibitem{letelier1977gauge}
P.S. Letelier, Phys. Rev. D \textbf{15}(4), 1055 (1977)

\bibitem{lund1976unified}
F.~Lund, T.~Regge, Phys. Rev. D \textbf{14}(6), 1524 (1976)

\bibitem{letelier1979clouds}
P.S. Letelier, Phys. Rev. D \textbf{20}(6), 1294 (1979)

\bibitem{ghosh2014cloud}
S.G. Ghosh, S.D. Maharaj, Phys. Rev. D \textbf{89}(8), 084,027 (2014)

\bibitem{belhaj2022shadows}
A.~Belhaj, Y.~Sekhmani, Gen. Relativ. Gravit. \textbf{54}(2), 17 (2022)

\bibitem{singh2020clouds}
D.V. Singh, S.G. Ghosh, S.D. Maharaj, Phys. Dark Universe \textbf{30}, 100,730
  (2020)

\bibitem{richarte2008traversable}
M.G. Richarte, C.~Simeone, Int. J. Mod. Phys. D \textbf{17}(08), 1179--1196
  (2008)

\bibitem{gogoi2023tideless}
D.J. Gogoi, U.D. Goswami, J. Cosmol. Astropart. Phys. \textbf{2023}(02), 027
  (2023)

\bibitem{nojiri2008modified}
S.~Nojiri, S.D. Odintsov, Phys. Rev. D \textbf{77}(2), 026,007 (2008)

\bibitem{capozziello2019extended}
S.~Capozziello, R.~D’Agostino, O.~Luongo, Int. J. Mod. Phys. D
  \textbf{28}(10), 1930,016 (2019)

\bibitem{gruber2014cosmographic}
C.~Gruber, O.~Luongo, Phys. Rev. D \textbf{89}(10), 103,506 (2014)

\bibitem{zhou2016new}
Y.N. Zhou, D.Z. Liu, X.B. Zou, H.~Wei, Eur. Phys. J. C \textbf{76}, 1--13
  (2016)

\bibitem{capozziello2021traversable}
S.~Capozziello, O.~Luongo, L.~Mauro, Eur. Phys. J. Plus \textbf{136}, 1--14
  (2021)

\bibitem{pade1892representation}
H.~Pad{\'e}, in \emph{Ann. Sci. de l'Ecole Norm. Superieure}, vol.~9 (1892),
  pp. 3--93

\bibitem{gracca2018cloud}
J.M. Gra{\c{c}}a, I.P. Lobo, I.G. Salako, Chin. Phys. C \textbf{42}(6), 063,105
  (2018)

\bibitem{mertens2014relevance}
T.G. Mertens, H.~Verschelde, V.I. Zakharov, J. High Energy Phys
  \textbf{2014}(11), 1--35 (2014)

\bibitem{curiel2017primer}
E.~Curiel, in \emph{Towards a theory of spacetime theories} (Springer, New
  York, NY, 2017), pp. 43--104

\bibitem{tolman1987relativity}
R.C. Tolman, \emph{Relativity, thermodynamics, and cosmology} (The Clarendon
  Press, Oxford, London, 1934)

\bibitem{oppenheimer1939massive}
R.J. Oppenheimer, G.M. Volkoff, Phys. Rev. \textbf{55}(4), 374 (1939)

\bibitem{gorini2008tolman}
V.~Gorini, U.~Moschella, A.Y. Kamenshchik, V.~Pasquier, A.A. Starobinsky, Phys.
  Rev. D \textbf{78}(6), 064,064 (2008)

\bibitem{ponce1993limiting}
J.P. de~Leon, Gen. Relativ. Gravit. \textbf{25}(11), 1123--1137 (1993)

\bibitem{visser2003traversable}
M.~Visser, S.~Kar, N.~Dadhich, Phys. Rev. Lett. \textbf{90}(20), 201,102 (2003)

\bibitem{kar2004quantifying}
S.~Kar, N.~Dadhich, M.~Visser, Pramana \textbf{63}(4), 859--864 (2004)

\bibitem{lobo2013new}
F.S.N. Lobo, F.~Parsaei, N.~Riazi, Phys. Rev. D \textbf{87}(8), 084,030 (2013)

\bibitem{penrose2002golden}
R.~Penrose, Gen. Relativ. Gravit. \textbf{34}(7), 1141--1165 (2002)

\bibitem{izmailov2019can}
R.N. Izmailov, A.~Bhattacharya, E.R. Zhdanov, A.A. Potapov, K.K. Nandi, Eur.
  Phys. J. Plus \textbf{134}(8), 384 (2019)

\bibitem{nandi2017ring}
K.K. Nandi, R.N. Izmailov, A.A. Yanbekov, A.A. Shayakhmetov, Phys. Rev. D
  \textbf{95}(10), 104,011 (2017)
\end{thebibliography}
\end{document}